\documentclass[10pt,conference]{apmc2022}
\ifCLASSINFOpdf
  \usepackage[pdftex]{graphicx}
\else
  \usepackage[dvips]{graphicx}
\fi
\usepackage{amsmath}

\usepackage{import}
\usepackage{color}
\usepackage{float}

\usepackage{tikz}

\newcommand\copyrighttext{%
	\footnotesize \textcopyright 2022 IEEE. Personal use of this material is permitted.  Permission from IEEE must be obtained for all other uses, in any current or future media, including reprinting/republishing this material for advertising or promotional purposes, creating new collective works, for resale or redistribution to servers or lists, or reuse of any copyrighted component of this work in other works. DOI: 10.23919/APMC55665.2022.9999980}
\newcommand\copyrightnotice{%
	\begin{tikzpicture}[remember picture,overlay]
	\node[anchor=south,yshift=10pt] at (current page.south) {\fbox{\parbox{\dimexpr\textwidth-\fboxsep-\fboxrule\relax}{\copyrighttext}}};
	\end{tikzpicture}%
}

\begin{document}
%
\title{Radar Classification of Vehicles Using a Ground-Reflection Model}



%
\author{\IEEEauthorblockN{Sören Kohnert\IEEEauthorrefmark{1},
Dominik Zoeke\IEEEauthorrefmark{2},
Reinhard Stolle\IEEEauthorrefmark{1}}
\IEEEauthorblockA{\IEEEauthorrefmark{1}
{\it HSA\_ired, University of Applied Science Augsburg, Germany}\\
soeren.kohnert@hs-augsburg.de}
\IEEEauthorblockA{\IEEEauthorrefmark{2}
{\it Siemens Mobility GmbH, Germany}}}


\maketitle
\copyrightnotice

\begin{abstract}
Classification of road users is important for traffic monitoring. The usability of a height estimate based on the two-ray ground-reflection model as a feature for the classification of vehicles is analyzed in this paper.
The four-ray ground-reflection model for fast chirp ramp sequence waveforms of FMCW radars is derived and simplified to the well-known two-ray ground-reflection model.
A spectrum from which the height of a target can be derived is obtained using the Lomb-Scargle periodogram. Measurements with two vehicle classes illustrate the approach and show that the model could be used as a feature to distinguish vehicles based on their height.
\end{abstract}


%
\IEEEpeerreviewmaketitle

\section{Introduction}
Radar sensors can provide evidence as a base for decisions in Smart Cities. Roadside perception gathers information e.g. to extend the view of automated vehicles \cite{masi:2021}. 
Such sensors should not only detect and localize objects but also classify them. 
Classification is done on features extracted from the measurement data.
The more independent features are, the more accurate the classification.

The two-ray ground-reflection is such a feature and it is well understood in communications and continuous wave (CW) radar \cite{etinger:2017}. The interference pattern emerges from the vector addition of the direct and reflected waves. 
The pattern 
allows the estimation of the height of a target without actual resolution in elevation. Further approaches to estimate the height of a target using a single antenna element exist, e.g. \cite{laribi:2017}, \cite{oh:2018}.
In an automotive radar context, the effect was first modeled by Bühren \cite{buehren:2007}. Diewald \cite{diewald:2011} uses the effect to distinguish between bridges and (stationary) vehicles at far distances. He uses interpolation to sample the distances uniformly and uses a high and low pass filter combination for the classification.

In contrast to the aforementioned applications, this paper adds the following contributions:
\begin{itemize}
    \item Derivation of the four-ray model for fast chirp ramp sequence waveforms \cite{gamba:2020} to describe the propagation factor.
    \item Sampling of the interference pattern, not only over the measurement cycles but over the chirps.
    \item Use of Lomb-Scargle periodogram \cite{vanderplas:2018} to obtain the height spectrum.
    \item Measurements of a car and truck using a stationary roadside mounted sensor.
\end{itemize}
The aim of this paper is not to implement a full classification system, but investigate if the features extracted from this effect make good classifiers for different types of vehicles.
\section{Ground-Reflection Model}
A point scatterer at a certain height approximates the millimeter wave reflections on the back and front of a car reasonably well \cite{buehren:2007}.
\vspace{-0.3cm}
\begin{figure}[H]
\centering
\begingroup%
  \makeatletter%
  \providecommand\color[2][]{%
    \errmessage{(Inkscape) Color is used for the text in Inkscape, but the package 'color.sty' is not loaded}%
    \renewcommand\color[2][]{}%
  }%
  \providecommand\transparent[1]{%
    \errmessage{(Inkscape) Transparency is used (non-zero) for the text in Inkscape, but the package 'transparent.sty' is not loaded}%
    \renewcommand\transparent[1]{}%
  }%
  \providecommand\rotatebox[2]{#2}%
  \newcommand*\fsize{\dimexpr\f@size pt\relax}%
  \newcommand*\lineheight[1]{\fontsize{\fsize}{#1\fsize}\selectfont}%
  \ifx\svgwidth\undefined%
    \setlength{\unitlength}{223.07751774bp}%
    \ifx\svgscale\undefined%
      \relax%
    \else%
      \setlength{\unitlength}{\unitlength * \real{\svgscale}}%
    \fi%
  \else%
    \setlength{\unitlength}{\svgwidth}%
  \fi%
  \global\let\svgwidth\undefined%
  \global\let\svgscale\undefined%
  \makeatother%
  \begin{picture}(1,0.40573013)%
    \lineheight{1}%
    \setlength\tabcolsep{0pt}%
    \put(0,0){\includegraphics[width=\unitlength,page=1]{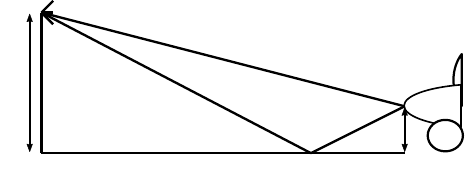}}%
    \put(-0.00283674,0.19792577){\color[rgb]{0,0,0}\makebox(0,0)[lt]{\lineheight{1.25}\smash{\begin{tabular}[t]{l}$h_S$\end{tabular}}}}%
    \put(0.80729211,0.09740551){\color[rgb]{0,0,0}\makebox(0,0)[lt]{\lineheight{1.25}\smash{\begin{tabular}[t]{l}$h_T$\end{tabular}}}}%
    \put(0.47460229,0.29250639){\color[rgb]{0,0,0}\makebox(0,0)[lt]{\lineheight{1.25}\smash{\begin{tabular}[t]{l}$d_d$\end{tabular}}}}%
    \put(0.63774702,0.10399934){\color[rgb]{0,0,0}\makebox(0,0)[lt]{\lineheight{1.25}\smash{\begin{tabular}[t]{l}$d_i$\end{tabular}}}}%
    \put(0,0){\includegraphics[width=\unitlength,page=2]{figures/signal_model.pdf}}%
    \put(0.45578479,0.00409751){\color[rgb]{0,0,0}\makebox(0,0)[lt]{\lineheight{1.25}\smash{\begin{tabular}[t]{l}$d$\end{tabular}}}}%
    \put(0.38344721,0.11679728){\color[rgb]{0,0,0}\makebox(0,0)[lt]{\lineheight{1.25}\smash{\begin{tabular}[t]{l}$\rho e^{j\varphi}$\\\end{tabular}}}}%
    \put(0,0){\includegraphics[width=\unitlength,page=3]{figures/signal_model.pdf}}%
  \end{picture}%
\endgroup%

\vspace{-0.2cm}
\caption{Geometry of the four-ray ground-reflection model for the interference pattern. On the left side, the stationary radar sensor is located on the pole at height $h_S$. On the right side, the vehicle is located with its main scattering center at height $h_T$.}
\label{fig:geometry}
\end{figure}
The four-ray ground-reflection model is typically used to describe these scenarios \cite{etinger:2017}.
For this purpose, Kerr \cite{kerr:1951} extended the well-known radar equation by the propagation factor $F_p$ to
\vspace{-0.3cm}
\begin{equation}
\begin{split}
P_r & = \frac{P_s G^2 \lambda^2 \sigma}{(4\pi)^3 r^4} \cdot |F_p|^2\\
\end{split}
\label{eq:radarEquation}
\end{equation}
with transmit power $P_s$, antenna gain $G$, wavelength $\lambda$, RCS $\sigma$, and distance $r$.

\subsection{Four-Ray Model for Fast Chirp Ramp Sequence Waveforms}
The complex voltage propagation factor $F_p$ is the ratio between the corresponding field strength when only free space effects are
considered $E_0$ and the four-way received electric field strength $E$.
\begin{equation}
F_p = \frac{|E|}{|E_0|} = \frac{|E_{dd}+2\cdot E_{di}+E_{ii}|}{|E_{dd}|} .
\label{eq:PropagatingFactorEField_4way}
\end{equation}
The indices represent the paths. $d$ stands for direct path, $i$ stands for indirect path.
Frequency Modulated Continous Wave (FMCW) radar sensors transmit a continuous carrier modulated by a ramp function
\begin{equation}
s^{tx}(t) = \exp{\left( j2\pi f_s t + j\pi\alpha t_s^2 \right)}.
\label{eq:SignalSend}
\end{equation}
$f_s$ is the frequency at the start of the ramp, $\alpha$ the sweep slope of the ramp and $t_s$ the time from the beginning of each ramp.
The time-delayed receive signal is proportional to
\begin{equation}
s^{rx}(t) = \left( \frac{2}{\tau(t) c_0}\right)^2 \exp{\left( j2\pi f_s (t-\tau(t)) + j\pi\alpha (t_s-\tau(t))^2 \right)} .
\label{eq:SignalReceived}
\end{equation}
The reflection coefficient of the target is neglected since it cancels out later anyway.
The time delay $\tau(t)$ depends on the distance of the target $r(t)$ and the speed of light $c_0$.
\begin{equation}
\tau(t) = \frac{r_{\mathrm{waythere}}(t) + r_{\mathrm {wayback}}(t)}{c_0}.
\label{eq:TimeDelay}
\end{equation}
The distance itself is time-dependent. For the moment, the time dependency will be neglected.
Inserting the distances of the different paths in Fig. \ref{fig:geometry} into \eqref{eq:SignalReceived}, and combining this with \eqref{eq:PropagatingFactorEField_4way} yields
\begin{equation}
\begin{split}
F_p = & \left| 1 + \frac{ 2 \cdot s^{rx}_{id}(t) }{s^{rx}_{dd}(t)}+ \frac{ s^{rx}_{ii}(t) }{s^{rx}_{dd}(t)} \right|\\
= \bigg| 1 &+ \frac{4d_d^2}{(d_i+d_d)^2} \cdot a \cdot \exp \left( j\frac{\pi\alpha}{c_0^2}(d_i^2+2d_i d_d -3d_d^2) \right)  \\
 &+ \frac{d_d^2}{d_i^2} \cdot a^2 \cdot \exp \left(j\frac{\pi\alpha}{c_0^2}4(d_i^2-d_d^2) \right) \bigg|
\end{split}
\label{eq:PropagationFactorFmcw}
\end{equation}
with
\vspace{-0.3cm}
\begin{equation}
a = \exp \left( -j\frac{2\pi}{c_0}(d_i - d_d) (f_s+\alpha t_s) + j\varphi \right) \cdot \rho,
\label{eq:a}
\end{equation}
where $\rho \cdot \exp(j\varphi)$ is the complex ground-reflection coefficient.
The distances of the indirect and direct path are
\begin{equation}
d_d = \sqrt{d^2 + (h_S-h_T)^2},   \hspace{0.5cm}   d_i = \sqrt{d^2 + (h_S+h_T)^2}. \\
\label{eq:distances}
\end{equation}
\subsection{Simplified Model}
The last exponential in the second and third term of \eqref{eq:PropagationFactorFmcw} can be neglected for typical sweep slopes ($<20 \frac{\mathrm{MHz}}{\mathrm{\mu s}}$ \cite{winner:2015}), since it is small compared to the other exponentials and causes only a constant phase shift in first-order approximation.
The difference between the distances of the paths can be approximated using the Laurent series to
\begin{equation}
d_i-d_d \approx \frac{2h_T h_S}{d}.
\label{eq:approxDiffDistances}
\end{equation}
This approximation is good if $d \gg h_S,h_T$ \cite{mahafza:2013}. 

Additionally, the attenuation can be ignored since $d_d \approx d_i$. The complex ground-reflection coefficient affects the phase and amplitude of $F_p$ but not its frequency which is used for classification. Hence, the coefficient is simplified to $-1$.
The magnitude of the propagation factor \eqref{eq:PropagationFactorFmcw} gets
\begin{equation}
\begin{split}
F_p &\approx 4 \sin^2 \left( \frac{\pi}{c_0}f_c\frac{2h_T h_S}{d} \right).
\end{split}
\label{eq:propagation_factor_approx}
\end{equation}
The last step leads from an FMCW specific equation to the well-understood two-ray ground-reflection model \cite{mahafza:2013}.
To get there, a centre frequency $f_c \approx f_s+\alpha t_s$ can be assumed since the maximum error (at $\alpha t_s = 0$ and $\alpha t_s = \mathrm{sweep\,bandwidth}$) for a 77GHz automotive radar is typically below $1\%$.
\section{Algorithm}
The propagation factor is amplitude modulated on the receive signal. 
To determine the height, the equation has to be solved for the frequency of the amplitude modulated receive signal. With a target (or the sensor) moving at constant velocity, a periodic sinusoid can be extracted that is sampled with $1/\Delta d$ distance between the samples.
After clearing the $\sin^2$ \eqref{eq:propagation_factor_approx} of its DC and normalizing it by the receive power to remove the influence of the RCS, the relation between spatial and height domain can be formulated using the Lomb-Scargle periodogram.
The Lomb-Scargle periodogram is identical to the result obtained by fitting a simple sinusoid in a least-squares sense to the data at each frequency and constructing a "periodogram" out of the goodness of fit \cite{vanderplas:2018}.
For the application shown here, it is necessary to calculate the bins up to the height of typical road participants, only.
The frequency axis of the periodogram can be scaled to the height of the target using
\vspace{-0.2cm}
\begin{equation}
\widehat{h_T}[k] = \frac{ \Delta_f}{2 h_S} \frac{c_0}{f_c} k.
\label{eq:target_height}
\end{equation}

A typical FMCW automotive sensor yields multiple ways to sample the pattern:
\begin{itemize}
    \item Sampling over all samples of a chirp ($f_s \approx 10\mathrm{MHz}$) 
    \item Sampling over the range spectra ($f_s \approx 20\mathrm{kHz}$)  
    \item Sampling over range-Doppler spectra ($f_s \approx 20\mathrm{Hz}$) 
\end{itemize}
Sampling over the range spectra (S.o.R.) allows for target separation in range. Sampling over the range-Doppler spectra (S.o.R.D.) allows for target separation in range and velocity.

\section{Measurements}
Sampling over all samples of a chirp is feasible in single target scenarios, only, which is unfeasible for roadside perception. The S.o.R. can separate targets in range. The number of sequential chirps is the number of samples taken from the interference pattern. The frequency can be influenced by adjusting the sensor height or the wavelength. The configuration chosen here would allow for nonaliased spectra even in the case of uniform sampling.
\subsection{Setup}
Height estimation by S.o.R. and S.o.R.D. is tested in a controlled environment to create free-field conditions with a single reflective surface, allowing to use the four-ray ground-reflection model derived above. The radar can be configured to different waveforms. Table \ref{tab:radPara} holds the parameters optimized for both measurement methods. 
\vspace{-0.3cm}
\renewcommand{\arraystretch}{1.1}
\begin{table}[H]
	\caption{Main parameters used for FMCW radar.}
	\small
	\centering
	\begin{tabular}{l|l|l|l}
		Parameter & S.o.R. & S.o.R.D. & Unit \\ \hline
		Centre frequency & $76.5$ & $76.5$ & $\mathrm{GHz}$ \\
		Sampling frequency & $4.17$ & $4.17$ & $\mathrm{MHz}$ \\
		Bandwidth & $342$  & $444.4$ & $\mathrm{MHz}$ \\
		Chirp duration & $61.4$ & $245.8$ & $\mathrm{\mu s}$ \\
		Pulse repetition frequency & $7.23$ & $3.39$ & $\mathrm{kHz}$\\
		Number of ramps & $512$ & $64$ & \\
		Maximum range & $56.0$ & $172.7$ & $\mathrm{m}$ \\
		Measurement cycle & $9$ & $18$ & $\mathrm{Hz}$ \\
		Sensor height & $2.63$ & $1$ & $\mathrm{m}$ \\
	\end{tabular}
\label{tab:radPara}
\end{table}
\renewcommand{\arraystretch}{1.0}	
\vspace{-0.3cm}
The signal processing scheme to preprocess the data consists of a classic FFT and CFAR approach \cite{gamba:2020}.
\subsection{Measurements on Vehicles}
The test vehicles drive with a constant velocity of 50km/h.
The test vehicles are a Mercedes Benz Atego 818 and a BMW 740li. The truck has a height of 3.45m and the car of 1.5m.
The height of the main reflection point on passenger cars can be expected to be in the range of 0.3m to 1m for common vehicle models \cite{buehren:2007}. For trucks, the main reflection point is expected to be higher or not a distinct reflection point at all. Both vehicles performed two runs. The vehicles span multiple range resolution cells. The closest local maximum to the sensor in the range domain is analyzed. 
\vspace{-0.3cm}
\begin{figure}[h]
\centering
\includegraphics[width=70mm]{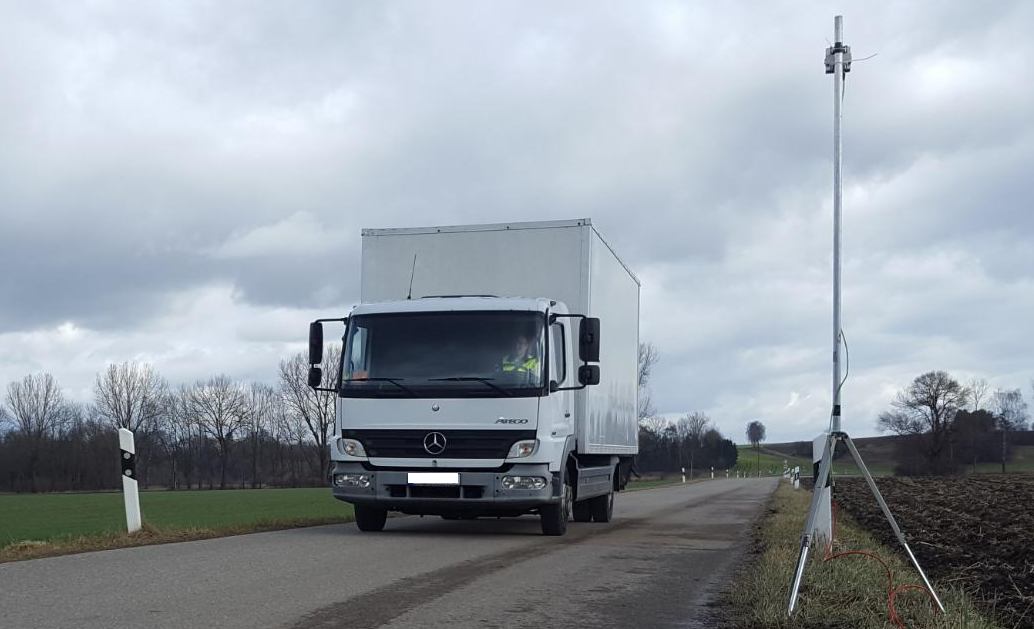}
\vspace{-0.2cm}
\caption{Measurement setup: on the left side one of the test vehicles, on the right side the sensor in the 2.63m configuration.}
\label{fig:messplatz}
\vspace{-0.1cm}
\end{figure}

\begin{figure}[h]
\vspace{-0.3cm}
\centering
\includegraphics[trim=5 0 15 12, clip,width=86mm]{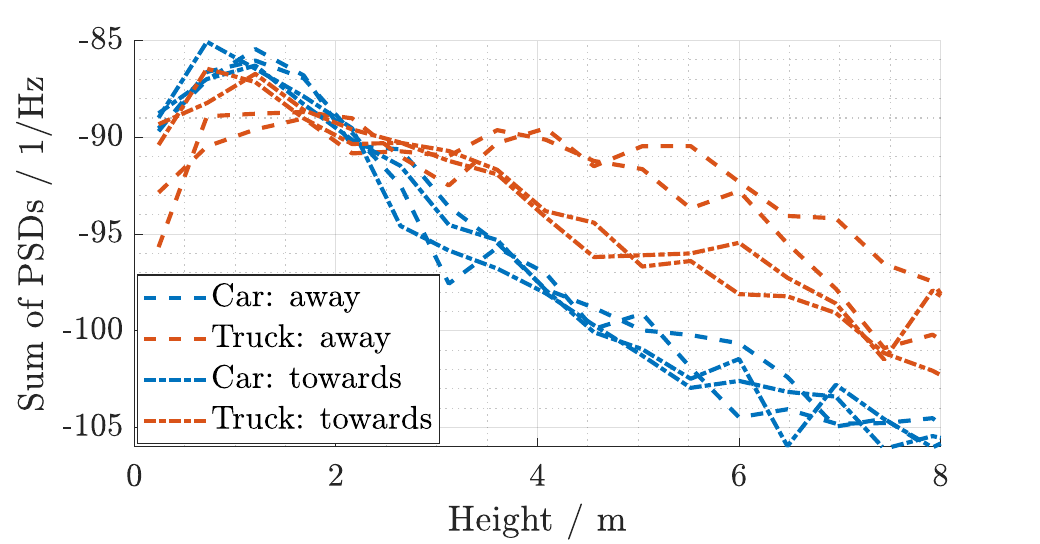}
\vspace{-0.2cm}
\caption{Sum of normalized PSDs of height estimates using S.o.R. measurements of both vehicles (car and truck) driving away from and towards to the sensor. Measurements from distances of 30m onward were taken into account. The spectra differ most at greater heights.}
\label{fig:meas_vehicles_short_time}
\vspace{-0.1cm}
\end{figure}

Fig. \ref{fig:meas_vehicles_short_time} shows the sum of all of the power spectral densities (PSDs) that are accumulated during a passage of the vehicle through the sensor`s field of view from 30m upwards obtained by S.o.R. The car has similar results regardless of the driving direction. The truck has greater height components, in particular when the back of the truck is visible. The back of the truck is higher than the driver's cabin in the front.

\begin{figure}[h]
\vspace{-0.2cm}
\centering
\includegraphics[trim=0 0 15 7, clip,width=86mm]{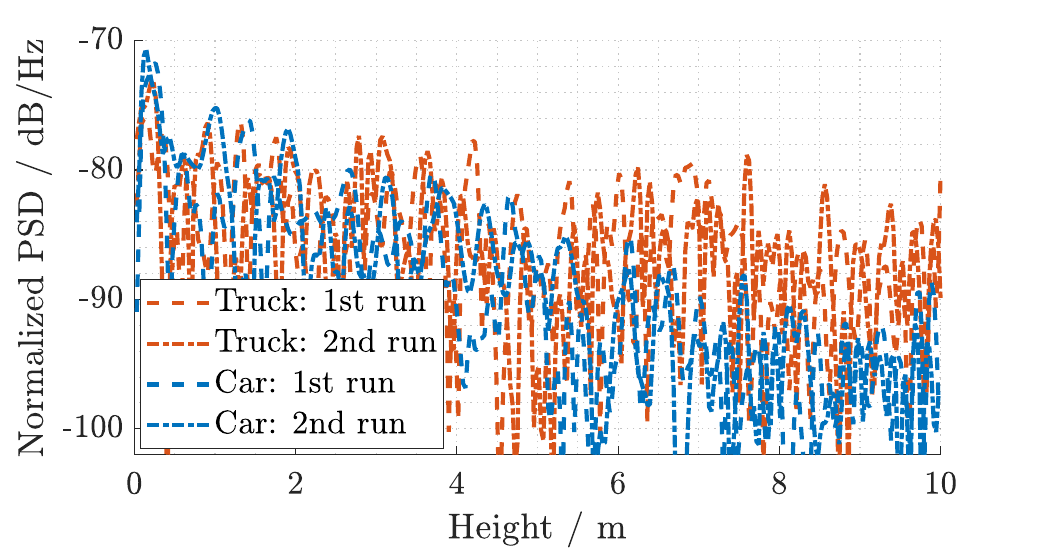}
\vspace{-0.2cm}
\caption{Normalized height PSDs obtained by S.o.R.D. Using measurements of the vehicle driving towards the sensor between 40m and 120m.}
\label{fig:meas_vehicles_long_time}
\vspace{-0.3cm}
\end{figure}

Fig. \ref{fig:meas_vehicles_long_time} depicts the PSDs gained by S.o.R.D. for both vehicles approaching the sensor. More power is concentrated at greater heights for the truck than for the car. The ratio for the power above e.g. 6m between the vehicle classes is 5.5 in the first run and 6.2 in the second. For the vehicles with the driving direction away from the sensor (not shown here), it is 3.7 and 6.8, respectively.
\vspace{-0.05cm}
\section{Conclusion}
\vspace{-0.05cm}
The modeling of the reflection of vehicles using the four-ray ground-reflection model for the purpose of classification seems feasible. Both sampling approaches showed potential to be used as a feature for classification.

One difficulty is to choose always the same range bin with regard to the coordinate system of the vehicle for the analysis.
In particular, this holds for distant objects where the dips in the reflected power approach the noise floor. 

The height spectra of the S.o.R. and S.o.R.D. both show frequency components beyond the expected height of the vehicles. 
The vector addition of more than one path causes additional frequencies in the height spectrum beyond those according to \eqref{eq:propagation_factor_approx}. The effect is more present when S.o.R.D, an explanation for that could be the long observation period (and changing incident angles) of the vehicle, allowing more scattering centers to contribute to the measurement, compared to S.o.R.

The proposed methods should further be tested in more reflective environments using a variety of vehicle types. 

\vspace{-0.05cm}
\bibliographystyle{IEEEtran}

\bibliography{IEEEabrv,IEEEexample}

\end{document}